\documentstyle[aps,prl,twocolumn]{revtex}

\begin{document}

\vspace{0.5cm}

\begin{center}
{\LARGE {\bf Interfering resonances in a }}

{\LARGE {\bf quantum billiard}}
\vspace*{0.5cm}

{\large {\bf 
E.~Persson$^\dagger$, K.~Pichugin$^*$, I.~Rotter$^\dagger$ and 
P.~${\rm \check S}$eba$^\ddagger$
}}\\[0pt]

\vspace{0.2cm}

{\it 
$^\dagger$Max-Planck-Institut f\"ur Physik komplexer Systeme, D-01187
Dresden, Germany, {\rm and}\\[0pt]
Technische Universit\"at Dresden, Institut f\"ur Theoretische Physik,
D-01062 Dresden, Germany \\[0pt]
$^*$Institute of Physics, Academy of Sciences\\
660036 Krasnoyarsk, Russia\\[0pt]
$^\ddagger$Nuclear Physics Institute, Czech Academy of Sciences,\\[0pt]
250 68 Re${\rm \check z}$ near Prague, Czech Republic\\[0pt]
}

\vspace{0.2cm} {\small 
$^\dagger$persson, rotter@mpipks-dresden.mpg.de \\[-.1cm]
$^*$zeos@krascience.rssi.ru \\[-.1cm]
$^\ddagger$seba@kostelec.czcom.cz 
}
\end{center}

\vspace{.3cm}


\vspace{.3cm}

\begin{abstract}
We present a method for numerically obtaining the positions, widths and
wavefunctions of resonance states in a two dimensional billiard connected
to a
waveguide. For a rectangular billiard, we study the dynamics of three
resonance poles lying separated from the other ones. As a function of
increasing coupling strength between the waveguide and the billiard two of the
states become trapped while the width of the third one
continues to increase for all coupling strengths. This behavior of the
resonance poles is reflected in the time delay function which can be
studied experimentally.\\ 
\end{abstract}

\vspace*{0.2cm}

In present-day high-resolution experimental studies, 
the properties of individual 
resonance states can be investigated even when the
 level density is high. As an example, 
 nuclear states  have recently been identified and studied experimentally 
at very high excitation energy \cite{br}. 
  Collisional damping at such an 
  energy is the same as at low excitation energy (ground-state domain). This
experimental result, being in contradiction to the standard statistical 
theory of nuclear reactions, can be justified by taking into account 
the interferences between resonance states
 arising from their interaction via the continuum \cite{pero}.
In atoms, the coherent coupling of autoionizing states is studied
experimentally e.g. in \cite{atom}. The authors point to the numerous 
possible applications of such investigations. Thus, the interferences 
between individual resonance states play an important r\^ole and need 
to be considered in detail.

 Theoretically, such interferences 
 have  been studied  in different 
fields of physics, see e.g. \cite{trap,trap2,FySo}.
One important result is the phenomenon of
resonance trapping, which arises from the interference of resonance
states coupled to the same decay channel. As a function of a parameter 
controlling the coupling to the continuum, the widths of all states
increase
as long as the states are isolated. At a certain critical value of the 
parameter where the resonance states start to overlap each other, the
widths bifurcate: the width of one of the states increases further while
the width of the other ones {\it decrease}, see e.g. \cite{trap2}. In other 
words: one of the states aligns with the channel
and becomes short lived while the other ones
decouple from the channel and
become long-lived in spite of the strong
 coupling to the continuum. 

The phenomenon of resonance trapping is theoretically well established but
not proven directly up to now in experimental studies. In this letter  we
investigate  the behavior of three neighboring resonances in a
two-dimensional billiard connected to a single waveguide. 
As a function of the coupling between the resonator and the waveguide,
we calculate both the position of the corresponding resonance poles and the
Wigner-Smith time delay function. The time delay function, containing
unique information on the interference between the resonances states, 
can be studied experimentally.

The model used is as follows. We consider a two dimensional billiard
coupled to a waveguide (for instance a flat electromagnetic resonator).  We
have to solve the equation 
\begin{equation}
(-\nabla ^{2}+\lambda V)\Psi =E\Psi   
\label{eq:origeq}
\end{equation}
where V is a potential barrier between the billiard and the attached lead.
We use the Dirichlet boundary condition, $\Psi =0$,
on the border of the billiard and of the waveguide. The waveguide has a
width
$W$ and the wavefunction inside it has the asymptotic form 
\begin{equation}
\Psi =\left(e^{ikx}-R(E)e^{-ikx}\right)u(y) \; .
\label{eq:assymp}
\end{equation}
Here $u(y)$ is the transversal mode in the
waveguide, $k$ is the wave number and $R(E)$ is the reflection coefficient.
It holds $E=k^{2}+(\pi /W)^{2}$. We choose the energy of the incoming wave
so that only the first transversal mode in the lead is open, i.e.
\begin{equation}
u(y)=sin(\frac{\pi }{W}y)\;.
\end{equation}
Since $|R(E)|=1$, we write $R(E)=exp(i\Theta (E))$ with $\Theta(E)$ real. 
A measurable quantity derived from $R$ 
is the Wigner-Smith time delay function,
\begin{equation}
\tau _{w}=\frac{d\Theta }{dE}\;.
\end{equation}
$\tau_w$ is the time the wave spends inside the billiard \cite{FySo}. 

The energies and widths of the resonance states are given by the poles of
the function $R(E)$ analytically continued into the lower complex plane.
To find the poles we use the exterior complex scaling method \cite
{complex}. The general idea is to study the system after a scaling
transformation applied to the $x$ coordinate, see \cite{complex,exterior}: 
$x\rightarrow \widetilde{x}=g(x)$. The function $g$ is chosen as
\begin{equation}
g(x)=\left\{ 
\begin{array}{ll}
x & x\geq x_{0} \\ 
\theta f(x) & x<x_{0}
\end{array}
\right. 
\end{equation}
with $f(x)$  such that $g(x)$ is three times smoothly differentiable and
the inverse transformation $g^{-1}(\tilde{x})$ exists. The attached waveguide
extends from $-\infty$ parallel to the x-axis and we choose $x_0$ to be
localized inside it. The related transformation of the wavefunction reads
\begin{equation}
\Psi \left( x,y\right)\longrightarrow\frac{1}{\sqrt{g^{\prime }(\tilde x)}}
\tilde\Psi(\tilde x,y)
\; .
\label{transform}
\end{equation}
Using it the equation (\ref{eq:origeq}) becomes 
\begin{eqnarray}
\nonumber
\left( -\frac{\partial }{\partial \tilde x}\left( \frac{1}
{g^{\prime \, 2}}\frac{%
\partial }{\partial \tilde x}\right) -\frac{\partial ^{2}}
{\partial y^{2}}\right) \tilde\Psi (\tilde x,y)+\hspace*{1.7cm}\\
\left( \lambda V(\tilde x,y)+\frac{2g^{\prime }g^{\prime \prime
\prime }-5g^{\prime \prime }{}^{2}}{4g^{\prime }{}^{4}}\right) \tilde\Psi
(\tilde x,y)=E\tilde\Psi (\tilde x,y)  
\label{scaledeq}
\end{eqnarray}
For a real parameter $\theta $, this equation is fully equivalent to 
(\ref{eq:origeq}) since the transformation (\ref{transform}) is unitary. 
Moreover, the two equations are fully identical for $x>x_{0}$. Since 
$x_0$ lies
inside waveguide the shape of
the resonator is not changed by the transformation (\ref{transform}) which
only rescales a part of the x-axis related to the waveguide. Moreover,
since  the
waveguide is oriented parallel to the x-axis, the transformation does not
change the
boundary of the system. For $\theta $ complex, (\ref{transform}) ceases to
be unitary and the spectral properties of \ (\ref{eq:origeq}) and (\ref
{scaledeq}) are different. The continuous spectrum of (\ref{eq:origeq}) 
extends over  $\left\langle \left( \pi /W\right) ^{2},\infty
\right) $, whereas the continuous spectrum of (\ref{scaledeq}) is rotated
into the complex plane and is equal to 
\begin{eqnarray}
\cup_{n=1:\infty }\left\{
\left( n\pi /W\right) ^{2}+\theta ^{-2}\left\langle 0,\infty \right)
\right\} \; . 
\end{eqnarray}
This is a  union of half-lines representing the 
continuous spectrum starting out from the real axis at every threshold
energy $(n\pi/W)^2$ with an angle $-2 \; arg \; \theta$.
The rotated continuous spectrum uncovers additive complex eigenvalues 
of (\ref{scaledeq}), the positions of which are independent of $\theta$.
These eigenvalues coincide with the resonance poles \cite{complex,exterior}.

In the following we study the time delay $\tau_w$ and the resonance poles
of a rectangular billiard of size $\Delta x\times\Delta y=2\times 3.14$
connected to a single waveguide with width $W=0.6$. We choose $V$ as a
rectangular potential barrier with height $1$ located at $-0.3\leq x\leq 0$.
By changing the parameter $\lambda$
we can tune the coupling between the waveguide and the resonator.
We calculate $\tau_w$ by solving Eq.~(\ref{eq:origeq}) with the 
Dirichlet boundary condition $\Psi=0$, and the asymptotic boundary
condition
(\ref{eq:assymp}) imposed at $x=-13$. The resonance poles are found by the
method of exterior complex scaling described above using $x_0=-2$.

In Fig.~1 we show the calculated time delay $\tau_w $ as a function of 
$\lambda $ and energy (panel a) as well as the dependence of the
resonance poles on $\lambda$ (panel b). 
At large $\lambda $ (weak coupling to the waveguide) 
we see three isolated resonance states. As $\lambda $ decreases (the coupling
to the waveguide increases) the lifetimes of all three states
decrease. The states attract each other  in energy. As
the resonances start to overlap, two of them become trapped
while the third one gets short lived. At further decreased $\lambda$
the lifetimes of the two trapped resonance states {\it increase}.
The motion of the poles is reflected in the time delay function.
The lifetime of the short lived state is, at small $\lambda$,
so short that it practically disappears when plotting the time delay.
The numerical errors in the distances among the resonances are so small that
the details of energy attraction and resonance trapping are stable.

Using the method of complex scaling we can also study the
wavefunctions of the resonance states (Gamow states). The interference
between the resonances states leads to a mixing of their wave functions
with respect to the eigenfunctions of the resonator ($\lambda\to\infty$). We
illustrate this phenomenon in Fig.~2, where  the wavefunctions of the three
resonance states are shown for $\lambda =44, 23.5$ and $0$, i.e. under the
condition when (a) they are isolated, (b) very near to one another and (c)
two of them are trapped while one is short lived.  

For $\lambda \rightarrow 0$, the amplitude of the wavefunction related to the
resonance state which finally evolves into a
short lived state is very small inside the resonator. This corresponds to
a very small time delay, i.e. to a small probability of staying inside
the  billiard. The trapping of the two long lived resonance states occurs
in two different ways. Both wavefunctions correspond to (almost) pure bound
states of the billiard as long as $\lambda $ is large and become mixed when
their distance in the complex energy plane to the other two states becomes 
small. At still smaller
$\lambda$ one of the resonance states demixes and approaches again 
its pure shape. In accordance to that, its energy  for small $\lambda $
approaches that for  large $\lambda $. The wavefunction of the other
trapped resonance state remains however  mixed. Accordingly, its
position at small $\lambda$ differs from that for large $\lambda $ and its
wavefunction is mixed with that of the short lived resonance state.

The analysis presented in this letter shows that open billiards provide an
excellent possibility to study the dynamics of resonance poles in detail.
A realization can be achieved by means of 
flat microwave resonators
connected to a  waveguide where 
the coupling strength to the channel
can be varied by hand.
Such systems allow therefore to investigate directly the
formation of different time-scales
(resonance trapping) by tracing  the
corresponding time delay function. In particular,
it is possible to show  the contraintuitive result
that the lifetimes of certain resonance
states  increase with increasing coupling to the continuum.
 The results of such investigations will help in analysing 
 high-resolution experimental data    in various fields of physics.
\\[.2cm]

\noindent
 \parbox{3.2cm}
{\bf Acknowledgement:}~Valuable discussions with T.~Gorin, B.~Mirbach,
M.~M\"{u}ller, J.~N\"ockel and G.~Soff are gratefully acknowledged.
This work has been partially supported by DFG, SMWK, Czech Grant Agency GAAV
1048804  and by
the Foundation for Theoretical Physics in Slemeno, Czech Republic.

\vspace*{.4cm}

{\bf Figure 1}\newline
Contour and surface plot of 
$log(\tau_w)$ ({\bf 1.a}) for a rectangular billiard (for parameters see
text). The darker the plot is, the longer is the time delay.
The motion of the corresponding resonance poles with $\lambda$ ({\bf 1.b}).
The positions of the resonance poles for $\lambda=44$ are denoted  
by squares and for $\lambda=23.5$ and $\lambda=0$ with large dots.\\

{\bf Figure 2}\newline
The wavefunctions of three resonance states for $\lambda =44$ (case a)), 
$\lambda =23.5$ (case b)) and $\lambda = 0$ (case c)).
The energies of the states at $\lambda=44$ are 38.6 
(state ${\rm N^{\underline o}}$ 1), 38.8 (${\rm N^{\underline o}}$ 2)
and 39.8 (${\rm N^{\underline o}}$ 3). The location of the potential barrier 
is marked by a rectangular in the waveguide.\\

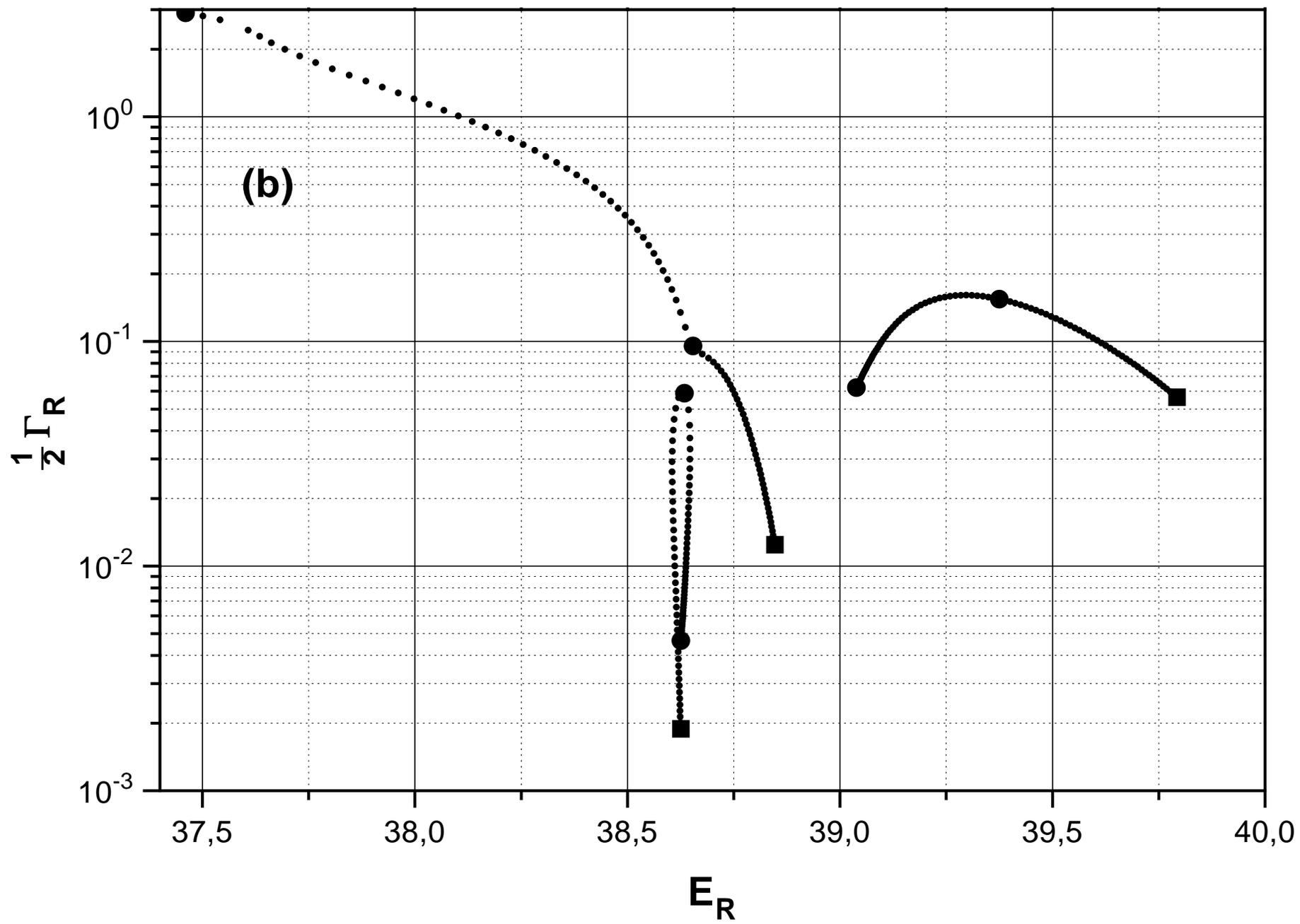
\end{document}